\documentclass{article}
\usepackage{spconf,amsmath,graphicx,diagbox}
\usepackage{pgfplots}
\usepackage{subfig}
\pgfplotsset{compat=newest}
\usetikzlibrary{plotmarks}
\usetikzlibrary{arrows.meta}
\usepgfplotslibrary{patchplots}

\newcommand{\betavar}{$\beta\text{-}\mathcal{VAR}$ }
\newcommand{\pdev}{$\mathcal{PDEV}$ }

\newcommand{\Figure}[1]{Figure~\ref{#1}}

\newcommand{\Equation}[1]{Equation~\eqref{#1}}

\newcommand{\Table}[1]{Table~\ref{#1}}

\newcommand{\Section}[1]{Section~\ref{#1}}

\def\imgs2{5.2 cm}
\title{An Improved Metric of Informational Masking for Perceptual Audio Quality Measurement}
\name{Pablo M. Delgado $^{\star}$, Jürgen Herre $^{\dagger \star}$ }

\address{$^{\star}$ International Audio Laboratories Erlangen $^{\ddagger}$, Am Wolfsmantel 33, 91058 Erlangen, Germany \protect\thanks{$^{\ddagger}$ A joint institution of the Friedrich-Alexander Universität Erlangen-Nürnberg (FAU) and Fraunhofer IIS Erlangen, Germany.} \\
$^{\dagger}$ Fraunhofer IIS, Am Wolfsmantel 33, 91058 Erlangen, Germany \\
Correspondence should be addressed to pablo.delgado@audiolabs-erlangen.de\\
}
\begin{document}
%
\maketitle
\begin{abstract}

Perceptual audio quality measurement systems algorithmically analyze the output of audio processing systems to estimate possible perceived quality degradation using perceptual models of human audition. In this manner, they save the time and resources associated with the design and execution of listening tests (LTs). Models of disturbance audibility predicting peripheral auditory masking in quality measurement systems have considerably increased subjective quality prediction performance of signals processed by perceptual audio codecs. Additionally, cognitive effects have also been known to regulate perceived distortion severity by influencing their salience. However, the performance gains due to cognitive effect models in quality measurement systems were inconsistent so far, particularly for music signals. Firstly, this paper presents an improved model of informational masking (IM) --- an important cognitive effect in quality perception ---  that considers disturbance information complexity around the masking threshold. Secondly, we incorporate the proposed IM metric into a quality measurement systems using a novel interaction analysis procedure between cognitive effects and distortion metrics. The procedure establishes interactions between cognitive effects and distortion metrics using LT data. The proposed IM metric is shown to outperform previously proposed IM metrics in a validation task against subjective quality scores from large and diverse LT databases. Particularly, the proposed system showed an increased quality prediction of music signals coded with bandwidth extension techniques, where other models frequently fail.

\end{abstract}
\section{Introduction}
\label{sec:intro}

Many of the widely-adopted and standardized Perceptual Audio Quality Measurement Systems (PQMS) for perceptually coded audio signals use models of human auditory perception \cite{PEAQ, POLQAcite}. Motivated by their success in perceptual audio codecs \cite{painter2000perceptual}, most of these methods have so far focused on models of peripheral auditory masking and disturbance loudness around and above the masking threshold using energetic considerations \cite{beerends1992a, biberger2018objective}. These metrics have been shown to strongly correlate with subjective degradation responses (i.e., grading annoyance or overall quality) in quality assessment tasks (see, e.g., \cite{PEAQ, MUSHRA}).

In addition to disturbance loudness, other cognitive and central audition aspects regulate perceived distortion severity. Newer systems incorporate, for example, modulation models for the evaluation of non-waveform preserving audio codecs \cite{Par2056_2019}. In \cite{beerends1996the}, Beerends proposed models for two important complementary cognitive phenomena in auditory perception: perceptual streaming (PS) and informational masking (IM) \cite{bregman1994auditory}. The PS model works under the assumption that added signal components (i.e., distortions) present in the signal under test ($SUT$) will normally fail to form a common percept with the input signal. As a separate percept, the distortion component will be generally judged as more unpleasant by the listening subjects, as if the distortion formed a single percept with the signal. On the other hand, missing signal components will be less objectionable, as they will integrate into one single percept with the input signal more easily. The cognitive effect of IM can be expressed as an increase of the masking threshold due to masker complexity. In \cite{beerends1996the}, the signal complexity is measured as the input signal's power deviation in time. The effects of PS and IM are thought to be complementary in that IM hinders the increase in perceived severity of a distortion due to PS. The model in \cite{beerends1996the} increased the prediction performance when applied to a disturbance loudness distortion metric. However, the performance gain in quality prediction of coded music signals was reported to be lower than in speech signals.

\section{Method}
\label{sec:method}

\begin{figure}[htb]
  \makebox[0.8\textwidth][l]{
      \includegraphics[width=0.48\textwidth]{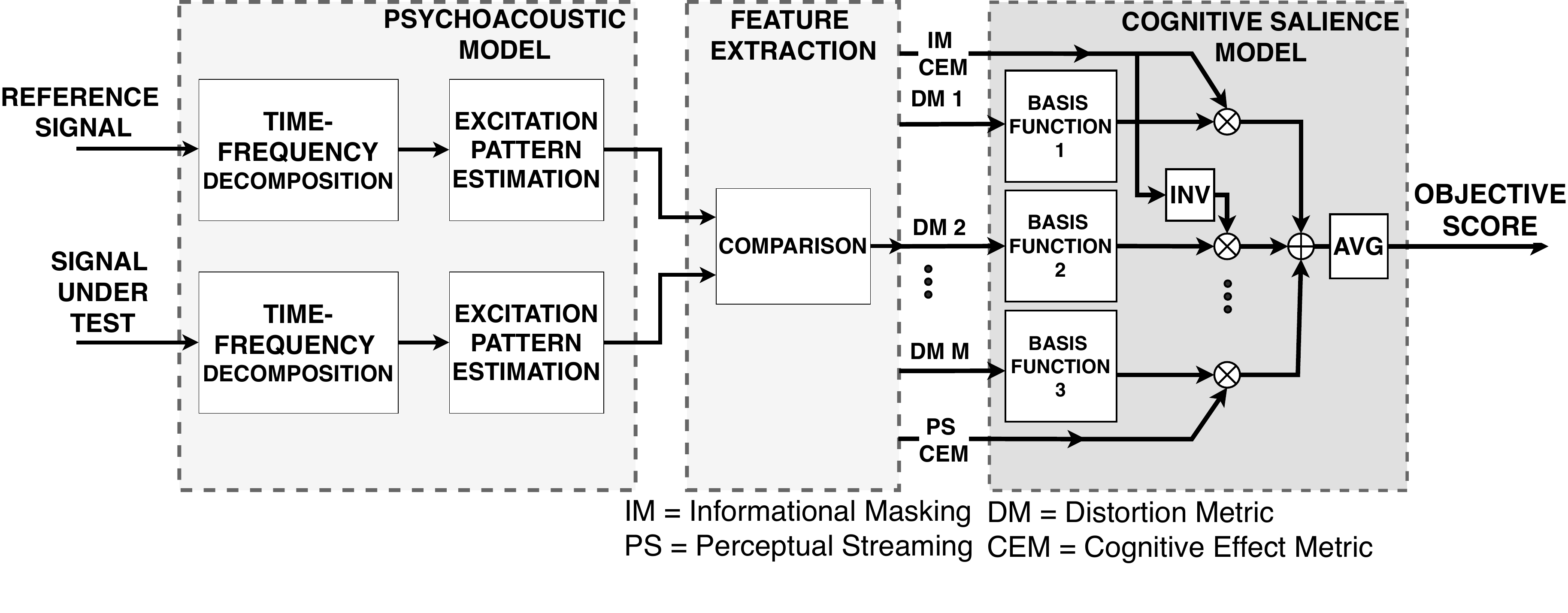}
  }
\caption{Block diagram of the proposed system (PEAQ-CSM) based on \cite{delgado2022data}.}
\label{fig:block_diagram}
\end{figure}

We base our work on selected modifications of a MATLAB implementation of the Perceptual Evaluation of Audio Quality method (PEAQ) \cite{PEAQ}, a widely-adopted and validated system. An unprocessed reference ($REF$) and different $SUT$ are compared in the auditory \textit{internal representation} domain. For the estimation of the internal representation, the time-frequency decomposition, excitation pattern estimation and comparison stages correspond to that of PEAQ's advanced version. The used distortion metrics (DM) are based on PEAQ's five Model Output Variables (MOVs) that describe different aspects of quality degradation, before time-frequency averaging. The Artificial Neural Network (ANN) of PEAQ that maps the MOVs to an overall quality score --- the mapping stage --- is replaced by the cognitive salience model proposed in \cite{delgado2022data} (\Figure{fig:block_diagram}). In the salience model, cognitive effect metrics (CEM) --- such as models of IM and PS ---,  interact with basis functions that map distortion metrics to the quality domain by weighting their contribution to an overall quality score. Only DM/CEM interactions that strongly predict distortion salience are included in the model. The salience model has been shown to outperform PEAQ's ANN mapping for the same number of input features. A time and frequency averaging over the whole duration of the analyzed signals is employed as a Basic Audio Quality (BAQ) \cite{MUSHRA} objective predictor. The resulting system is referred to as PEAQ-CSM.

The models of the three aspects of quality modelling involved in this work: perceived disturbance loudness, IM and PS are described in the following sections.

\subsection{Partial Loudness of Disturbances}
\label{sec:partial_loudness}
The perceived loudness of distortions has been shown to be an important predictor of audio quality degradation. The developers of PEAQ proposed a distortion metric in which disturbance loudness is not only modeled well above the masking threshold, but also in the vicinity of the masking threshold (often called \textit{partial loudness}). It is believed that the loudness behaviour of sound events in this region considerably differs from behaviors at well below or above the masking threshold \cite{moore1997model}. These models specify that the loudness of sounds in this region is not zero, but take progressively smaller values as they approach the threshold of audibility. Some objective quality metrics map the perceived disturbance severity in this region to detection probabilities \cite{PEAQ}. 

In PEAQ's disturbance loudness metric, the instantaneous values of the excitation patterns for $REF$ and $SUT$, $E_R(n,k)$ and $E_T(n,k)$ are compared at a time frame $n$ and auditory band $k$ \footnote{Indices $(n,k)$ are omitted in \Equation{eq:noiseloud} for clarity}:

\begin{align}
  \label{eq:noiseloud}
  \resizebox{0.43\textwidth}{!}{
  $N' = 0.068 \left( \frac{1}{s_{T}} \frac{E_{th}}{E_0} \right)^{\gamma} \left[ \left( 1+ \frac{max(s_{T}E_{T}-s_{R}E_{R},0)}{ E_{th} + s_{R}E_{R} \cdot \beta} \right)^{\gamma} - 1 \right]$
  }, \\
  \beta = e^{ -\alpha(E_{T}-E_{R})/E_{ref}}.
\end{align}
The coefficients $E_{th}$ and $E_0$ are scaling constants and $\gamma = 0.23$ \cite{zwicker1967ohr}. The terms $s_{R,T}$ incorporate the effect of signal modulations in the masking threshold level to account for asymmetry of masking effects between noise and tones \cite{PEAQ}. The term $\beta$ describes the disturbance behavior in the vicinity of the masking threshold and $\alpha = 1.5$ determines the amount of partial masking. 

\subsection{Previous Approach in Models of PS and IM}
\label{sec:prev_approach}

In order to analyze the previous approach in modelling PS and IM of \cite{beerends1996the}, we implement Barbedo and Lopes \cite{barbedo2005a} adaptation into the PEAQ framework. Considering the complementary effect between PS and IM mentioned in the Introduction, a disturbance loudness distortion metric such as the one of \Equation{eq:noiseloud} has been extended in \cite{beerends1992a} as follows:

\begin{align}
  \label{eq:IMPSold}
  N'_{IMPS}(n,k) =& \mathcal{IMPS}(n,k) \cdot N'(n,k). \\
  N'_{IMPS}(n,k) =& \frac{C \cdot \mathcal{PS}(n,k)^a \cdot N'(n,k)}{\mathcal{PDEV}(n)^b + C}
\end{align}
the constants $a,b$ and $C$ are chosen to directly interact with the values of a noise loudness metric (e.g., $N'$), whereas Lopes and Barbedo \cite{barbedo2005a} considered the term $\mathcal{IMPS}$ as a separate DM in the mapping stage. The term $\mathcal{PS}$ is a CEM estimating the amount of perceptual streaming, calculated as:

\begin{align}
  \mathcal{PS}_0(n,k) =& \frac{E_{T}(n,k)+1}{E_{R}(n,k)+1}, \label{eq:PS0}\\
  \mathcal{PS}(n,k) =& 0.5 \cdot \mathcal{PS}_0(n,k) + 0.5 \cdot \mathcal{PS}_0(n-1,k). \label{eq:PSFilter}
\end{align}
The weighted average of previous frames in \Equation{eq:PSFilter} increases the value of the CEM when distortions are present over a consecutive amount of frames. The PS CEM can interact with the values of a distortion metric in that PS can regulate the perceived severity of the distortion. Increases in PS are expected to reinforce the predicted degradation as measured by a distortion metric.

Similarly, an estimate of the amount of IM is implemented as a CEM: 
\hphantom{s}
\begin{align}
  \mathcal{PDEV}(n,k) =& | E_R(n,k) - \bar{p}(n,k) |  \\
  \mathcal{IM}(n) = \mathcal{PDEV}(n) =& \frac{1}{K} \sum_{1}^K \mathcal{PDEV}(n,k) \label{eq:pdev}
\end{align}
where $\bar{p}$ is the mean signal power calculated over a 20 ms time window. The CEM $\mathcal{PDEV}(n)$ predicts the amount of IM at a given time $n$. Disturbances above the masking threshold as calculated in \Equation{eq:noiseloud} can be masked when the power variation increases, as the signal becomes more complex. In this case, the metric may overestimate quality degradation by not considering audibility threshold increases due to increased signal complexity. Therefore, the $\mathcal{IM}(n)$ CEM can interact with it to counteract the overestimation.

\subsection{Proposed Approach}
\label{sec:prop_approach}
The previous approach in modeling IM using signal power variations considers the properties of the input signal independently of the operational region of masker signal and disturbances mentioned in \Section{sec:partial_loudness}. Additionally, previous approaches either specify that IM and PS only interact with a disturbance loudness metric (\cite{beerends1992a}, e.g., \Equation{eq:IMPSold}) or rely on a general-purpose regression procedure to find interactions of CEMs with other DMs (\cite{barbedo2005a}). Our proposed approach is related to these two aspects.

Firstly, we propose an IM model based on disturbance behavior in the partial loudness region using the following arguments: as the disturbance severity in this region can be mapped to a detection probability, the signals' stochastic properties (e.g., power deviations) will have a considerable influence in modelling quality degradation. Additionally, the majority of the IM models also consider masker variations as random variables modelling masking noise \cite{DurlachNotesInfo}. We therefore assume that a stochastic model of disturbance variation --- particularly in the partial loudness region --- will represent a more accurate picture of the intrinsic cognitive processes involved in IM than models considering overall signal power deviations in all loudness regions.

\begin{align}
  \mathcal{IM}(n) = \beta\text{-}\mathcal{VAR}(n) = \frac{1}{K} \sum_{1}^K var(\beta(n,k))
\end{align}
where $var(\beta)$ is the moving variance of the near-threshold masking term of \Equation{eq:noiseloud} with a time-window of 100 ms \footnote{Based on stimulus duration in \cite{lutfi1993model}} and a normalization factor of $N-1$ time samples.

Secondly, we propose a more complex interaction model of the IM CEM with the available DMs in PEAQ than the previous approaches. In the previous approach, the nature of the interactions of the cognitive effect metrics with quality metrics was based on theoretical considerations (i.e., PS and IM interact with disturbance loudness to influence perceived quality degradation) and it was validated by considering overall quality prediction performance on listening test DBs. In our approach, we do not assume a fixed interaction with a DM as in \Equation{eq:IMPSold}, but perform the CEM-DM salience interaction analysis that we presented in \cite{delgado2022data}. In the cited work, the interactions in the quality mapping model are chosen based on the Pearson correlation of the CEM values against a data-driven salience metric defined on multiple DMs (e.g., roughness, band limitation, and so on ...), over a representative listening test database. The CEM values that show a strong correlation with salience metric values calculated for a given DM are assumed to predict the salience of said DM's measured distortion. Only strong CEM-DM interactions stemming from the salience analysis are incorporated into PEAQ-CSM.

\section{Validation Experiment Design}
\label{sec:Experiment}

A system that uses the proposed IM and PS model (PEAQ-CSM $\beta\text{-}\mathcal{VAR}$) was validated using a series of extensive listening test DBs that contain carefully collected subjective data on a wide variety of audio-coding related signal degradations in different quality ranges. The DBs are the Unified Speech and Audio Coding Verification Tests 1 and 3 (USAC VT) \cite{USACdatabase} and the Enhanced Low Delay AAC Verification Test DBs (ELD VT) \cite{ELDdatabase}. The test signals include music (isolated instruments and ensembles), speech (single and interfering speakers) with different levels of reverberation and other critical signals such as applause recordings. In total, 639 mean quality scores pooled from thousands of individual listener responses in different labs were collected with the Multiple Stimuli with Hidden Reference and Anchor (MUSHRA) method \cite{MUSHRA} for each audio item and codec. A detailed description of the experimental methodology and analysis can be found in both documents \cite{USACdatabase,ELDdatabase}. Using these DBs, the performance of PEAQ-CSM \betavar was compared to a first baseline system PEAQ-CSM using the previous measure of IM (PEAQ-CSM $\mathcal{PDEV}$) and to a second baseline PEAQ-CSM that does not incorporate any effects of IM and PS. Additionally, two established objective quality metrics, PEAQ DI (Advanced Version) and ViSQOL Audio (NSIM) \cite{Visqol3} were also included in the comparison. The overall performance was measured in terms of MUSHRA Score/Objective score correlation $R$ with a third-order polynomial function pre-mapping, as recommended in \cite{EvalObjective}.

\section{Results and Discussion}

\begin{table}[t!]
  \centering
    \resizebox{\columnwidth}{!}{%
      \begin{tabular}{l|c|c|c}
        \diagbox[width=\dimexpr \textwidth/8+2\tabcolsep\relax, height=1cm]{\textbf{CEM}}{\textbf{DM}} & $RmsNoiseLoud$ & $SegmentalNMR$ & $EHS$ \\ \hline
        $\mathcal{PS}$                  & 0.5 & 0.4 & 0.73 \\
        $\mathcal{PDEV}$                & -0.44 & -0.67 & -0.60 \\
        $\beta\text{-}\mathcal{VAR}$ & -0.34 & -0.73 & -0.85 \\
        \multicolumn{4}{l}{\footnotesize{DM: distortion metric, CEM: cognitive effect metric, $\mathcal{PS}$: perceptual streaming, }} \\
        \multicolumn{4}{l}{\footnotesize{$\mathcal{PDEV}$: signal power deviation, $\beta\text{-}\mathcal{VAR}$: proposed method.}}        
      \end{tabular}
      }
				\caption{CEM-DM Salience correlation values for the analysis DB USAC VT1, the proposed CEMs and PEAQ's MOVs. All other CEM-DM salience correlation values are less than 0.5 and $CI_{95\%} < 0.01$.}
	\label{tab:CEMDMsalience}
\end{table}

The distortion metric/CEM salience interaction analysis results are shown in \Table{tab:CEMDMsalience}. It can be seen that the proposed $\beta\text{-}\mathcal{VAR}$ IM metric shows a moderate to strong prediction power over the perceptual salience of errors within harmonic structures, measured by the $EHS$ MOV, and the salience of coding noise above the masking threshold, measured by the $SegmentalNMR$ MOV \cite{PEAQ}. The negative correlation values for IM indicate that increasing CEM values predict decreasing salience i.e., a larger IM effect size decreases the perceived severity of the associated distortion, as discussed in \Section{sec:prev_approach}. Likewise, increasing CEM PS values predict the increased salience of the distortions measured by $EHS$. The selected interactions of the IM and PS models with the mentioned MOVs are included into PEAQ-CSM. Remarkably, there is a weak interaction of the IM and PS effect sizes with PEAQ's partial disturbance loudness metric, $RmseNoiseLoud$. For the data analyzed, an interaction as suggested by \Equation{eq:IMPSold} is not justified. This interaction is therefore not included in the PEAQ-CSM method.

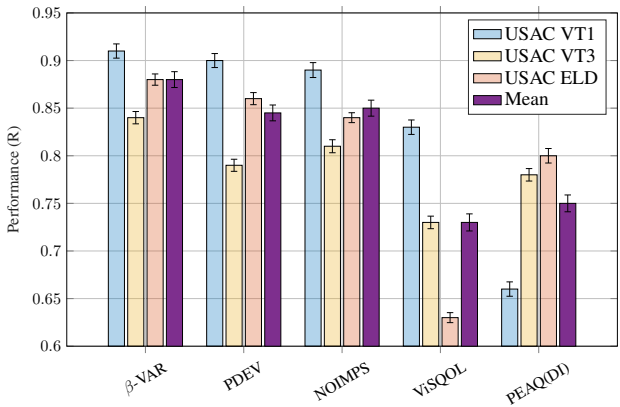
\begin{figure}[htb]
  \makebox[\textwidth][l]{
    \resizebox{0.61\width}{!}{

\definecolor{mycolor1}{rgb}{0.00000,0.44700,0.74100}%
\definecolor{mycolor2}{rgb}{0.85000,0.32500,0.09800}%
\definecolor{mycolor3}{rgb}{0.92900,0.69400,0.12500}%
\definecolor{mycolor4}{rgb}{0.49400,0.18400,0.55600}%

\pgfplotstableread{
x                       VT1   VT1ci+   VT3   VT3ci+   ELD     ELDci+   MEAN   MEANci+  
$\mathcal{\beta}$-VAR	0.91   0.00747  0.84  0.00647 0.88    0.00593   0.88   0.00832    
PDEV                    0.90   0.00735  0.79  0.00635 0.86    0.00635   0.845  0.00831    
NOIMPS                  0.89   0.00783  0.81  0.00683 0.84    0.00520   0.85   0.0084    
ViSQOL                  0.83   0.00759  0.73  0.00659 0.63    0.00525   0.73   0.00897    
PEAQ(DI)                0.66   0.00755  0.78  0.00655 0.8     0.00762   0.75   0.00884    

}{\mytable}

\begin{tikzpicture}

\begin{axis}[%
    ybar,
    width=12cm,
    height=7.269cm,
    at={(0cm,0cm)},
    scale only axis,
    bar shift auto,
    log origin=infty,
    symbolic x coords={$\mathcal{\beta}$-VAR, PDEV,NOIMPS,ViSQOL, PEAQ(DI)},
    enlarge x limits=0.2,
    xtick=data,
    xticklabel style={rotate=30},
    ymin=0.6,
    ymax=0.95,
    ylabel style={font=\color{white!15!black}},
    ylabel={Performance (R)},
    axis background/.style={fill=white},
    xmajorgrids,
    ymajorgrids,
    legend style={legend cell align=left, align=left, draw=white!15!black},
    legend style={font=\large},
    ]

\addplot[fill=mycolor1, draw=black, area legend, fill opacity=0.3, error bars/.cd, y dir=both, y explicit]
  table [x index=0,y=VT1, y error plus=VT1ci+, y error minus=VT1ci+] {\mytable};
\addlegendentry{$\text{USAC VT1}$}

\addplot[fill=mycolor3, draw=black, area legend, fill opacity=0.3, error bars/.cd, y dir=both, y explicit]
  table [x index=0,y=VT3, y error plus=VT3ci+, y error minus=VT3ci+] {\mytable};
\addlegendentry{$\text{USAC VT3}$}

\addplot[fill=mycolor2, draw=black, area legend, fill opacity=0.3, error bars/.cd, y dir=both, y explicit]
  table [x index=0,y=ELD, y error plus=ELDci+, y error minus=ELDci+] {\mytable};
\addlegendentry{$\text{USAC ELD}$}

\addplot[fill=mycolor4, draw=black, area legend, error bars/.cd, y dir=both, y explicit]
  table [x index=0,y=MEAN, y error plus=MEANci+, y error minus=MEANci+] {\mytable};
\addlegendentry{$\text{Mean}$}

\end{axis} 
\end{tikzpicture}
  }}
\caption{Overall system performance. Error bars denote 95\% confidence intervals ($CI_{95\%}$). }
\label{fig:overall_system_performance}
\end{figure}

The overall performance results are shown in \Figure{fig:overall_system_performance}. On average, the system that incorporates the proposed IM model (PEAQ-CSM $\mathcal{PDEV}$ ) significantly outperforms all other variants in cases (violet bar). Multiple comparison t-tests showed significant improved performance of the proposed method when compared to $\mathcal{PDEV}$ for the VT3 and ELD DBs. Significant differences were found in all 3 databases when compared to a system with no IM and PS model and to other established objective quality metrics. 

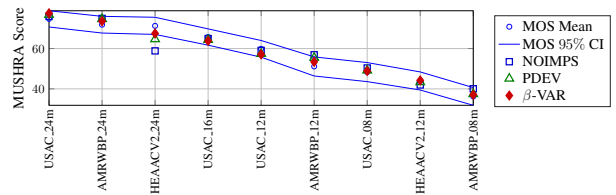
\begin{figure}[htb]
  \makebox[\textwidth][l]{
    \resizebox{0.47\width}{!}{
%
%
\definecolor{mycolor1}{rgb}{0.00000,0.44700,0.74100}%
\definecolor{mycolor2}{rgb}{0.00000,0.0000,0.80100}%
\definecolor{mycolor3}{rgb}{0.00000,0.50000,0.00100}%
\definecolor{mycolor4}{rgb}{0.80000,0.00000,0.00000}%
\begin{tikzpicture}
  \pgfplotsset{
  every axis legend/.append style={ at={(1.05,0.95)}, anchor=north west,legend columns = 1}}
\begin{axis}[%
width=12cm,
height=2.676cm,
at={(0cm,0cm)},
scale only axis,
unbounded coords=jump,
xmin=1,
xmax=9,
xtick={1,2,3,4,5,6,7,8,9},
xticklabels={{$\text{USAC\_24m}$},{$\text{AMRWBP\_24m}$},{$\text{HEAACV2\_24m}$},{$\text{USAC\_16m}$},{$\text{USAC\_12m}$},{$\text{AMRWBP\_12m}$},{$\text{USAC\_08m}$},{$\text{HEAACV2\_12m}$},{$\text{AMRWBP\_08m}$}},
xticklabel style={rotate=90},
ymin=31.8568342027717,
ymax=78.72956766118,
ylabel style={font=\large},
ylabel={$\text{MUSHRA Score}$},
axis background/.style={fill=white},
xmajorgrids,
ymajorgrids,
legend style={legend cell align=left, align=left, draw=white!15!black},
legend style={font=\large},
]
\addplot [color=blue, only marks, mark=o, mark options={solid, blue}]
  table[row sep=crcr]{%
1	74.7066656403941\\
2	71.820197044335\\
3	71.2483066502463\\
4	65.7104371921182\\
5	59.8473676108374\\
6	51.1239224137931\\
7	48.314039408867\\
8	43.9206434729064\\
9	36.2441502463054\\
};
\addlegendentry{MOS Mean}

\addplot [color=blue]
  table[row sep=crcr]{%
1	78.72956766118\\
2	75.9631665546323\\
3	75.4971418094238\\
4	69.7457954888819\\
5	63.9827875475055\\
6	55.8538034332008\\
7	53.0378404478732\\
8	48.4029752058733\\
9	40.6314662898392\\
};
\addlegendentry{MOS $95\%$ CI}

\addplot [color=blue, forget plot]
  table[row sep=crcr]{%
1	70.6837636196082\\
2	67.6772275340377\\
3	66.9994714910688\\
4	61.6750788953545\\
5	55.7119476741693\\
6	46.3940413943854\\
7	43.5902383698608\\
8	39.4383117399395\\
9	31.8568342027717\\
};






\addplot [color=mycolor2, line width=1.0pt, only marks, mark size=2.5pt, mark=square, mark options={solid, fill=mycolor2, mycolor2}]
  table[row sep=crcr]{%
  1	75.9630245736437\\
  2	74.9218520420239\\
  3	58.8398304658558\\
  4	64.9353751534986\\
  5	59.0476418494288\\
  6	56.8874851680706\\
  7	50.360588999338\\
  8	41.9425123648197\\
  9	40.037419063124\\
};
\addlegendentry{NOIMPS}


\addplot [color=mycolor3, line width=1.0pt, only marks, mark size=4 pt, mark=triangle, mark options={solid, fill=mycolor3, mycolor3}]
  table[row sep=crcr]{%
1	76.7301016281882\\
2	74.6295837519204\\
3	64.6224080621002\\
4	64.2808022104501\\
5	57.3869865238627\\
6	55.4395275098583\\
7	49.1269153600172\\
8	43.3143434127712\\
9	37.4050612206347\\
};
\addlegendentry{PDEV}



\addplot [color=mycolor4, line width=1.0pt, only marks, mark size=3.5pt, mark=diamond*, mark options={solid, fill=mycolor4, mycolor4}]
  table[row sep=crcr]{%
1	77.4781856283163\\
2	73.5559442607813\\
3	67.476479970732\\
4	63.7786533969514\\
5	57.1202285972435\\
6	53.3982294461906\\
7	48.8703796649362\\
8	44.1936965878172\\
9	37.0639321268344\\
};
\addlegendentry{$\mathcal{\beta}$-VAR}
\end{axis}


\end{tikzpicture}%
  }}
\caption{Pooled coding condition subjective quality scores and objective quality predictions (MUSIC ITEMS USAC VT 1).}
\label{fig:plot_CEM_vs_MOV}
\end{figure}

\begin{figure}[h!]
  \centering
  \subfloat[EHS]{
  \label{fig:EHS}
  \resizebox{0.27\columnwidth}{2.1cm}{
%
%
\begin{tikzpicture}

\begin{axis}[%
xtick=\empty,
width=10.348cm,
height=9cm,
at={(0cm,0cm)},
scale only axis,
point meta min=-0.000846092096828786,
point meta max=0.00317585906500741,
axis on top,
xmin=0.5,
xmax=101.5, 
y dir=reverse,
ymin=0.5,
ymax=40.5,
ytick={1,20,40},
xlabel style={font=\Huge},
xlabel={TIME},
yticklabels={{18 kHz},{3.2kHz},{0.05 kHz}},
yticklabel style={font=\Huge},
axis background/.style={fill=white},
legend style={font=\footnotesize},,
]
\addplot [forget plot] graphics [xmin=0.5, xmax=101.5, ymin=0.5, ymax=40.5] {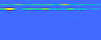};
\end{axis}
\end{tikzpicture}%
}
  }
  \qquad
  \subfloat[PDEV]{
  \label{fig:PDEV}
  \resizebox{0.20\columnwidth}{2.05cm}{
%
%
\begin{tikzpicture}

\begin{axis}[%
xtick=\empty,
ytick=\empty,
width=10.348cm,
height=9cm,
at={(0cm,0cm)},
scale only axis,
point meta min=1.79381611178364e-05,
point meta max=71176.9580515863,
axis on top,
xmin=0.5,
xmax=101.5,
y dir=reverse,
ymin=0.5,
ymax=40.5,
xlabel style={font=\Huge},
xlabel={TIME},
axis background/.style={fill=white},
legend style={font=\footnotesize},,
]
\addplot [forget plot] graphics [xmin=0.5, xmax=101.5, ymin=0.5, ymax=40.5] {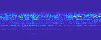};
\end{axis}
\end{tikzpicture}%
}
  }
  \qquad
  \subfloat[$\mathcal{\beta}$-VAR]{
  \label{fig:BVAR}
  \resizebox{0.20\columnwidth}{2.05cm}{
%
%
\begin{tikzpicture}
\begin{axis}[%
xtick=\empty,
ytick=\empty,
width=10.348cm,
height=9cm,
at={(0cm,0cm)},
scale only axis,
point meta min=0,
point meta max=0.0143073882343219,
axis on top,
xmin=0.5,
xmax=101.5,
y dir=reverse,
ymin=0.5,
ymax=40.5,
xlabel style={font=\Huge},
xlabel={TIME},
axis background/.style={fill=white},
legend style={font=\footnotesize},
]
\addplot [forget plot] graphics [xmin=0.5, xmax=101.5, ymin=0.5, ymax=40.5] {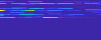};
\end{axis}
\end{tikzpicture}%
}
  }
\caption{T/F plots of the $EHS$ MOV and investigated CEMs for a violin recording coded with bandwidth extension techniques \cite{dick2017generation}.}
\label{fig:three_pane}
\end{figure}

One of the strong points of the proposed model is illustrated in \Figure{fig:plot_CEM_vs_MOV}. The proposed informational masking model reduces the overestimation in quality degradation for the coding condition $HEAACV2\_24m$ for music items. The HE-AAC v2 codec is known for its use of bandwidth extension techniques \cite{heaacv2}, which can generate coding artifacts with harmonic structures and disturbances related to small modulations of harmonic components in tonal signals \cite{ArtifactsinPAC} in the high frequency range. These artifacts are picked up by the $EHS$ MOV as expected (see \Figure{fig:EHS}). However, at the used bitrate (24 kbps, mono), subjective scores indicate that the disturbances are not perceived as severely as predicted by said MOV. The proposed IM metric \betavar counteracts the overestimation of $EHS$ in higher frequencies (\Figure{fig:BVAR}) by considering the IM effect caused by small disturbance power variations around the masking threshold. In contrast, the \pdev metric based on total signal power variation fails to identify the region where IM interacts with the perceived distortion (\Figure{fig:PDEV}). Limitations in the $EHS$ MOV for music signals were also reported in \cite{DelgadoPEAQ}.

\section{Conclusion and Future Work}

The consideration of models of central and cognitive audition in PQMSs can improve quality prediction performance. The proposed IM metric increased a PQMS performance in comparison to previous approaches \footnote{The authors would like to thank Aalok Gupta for his support in preliminary discussions and data analysis.}. This is assumed to be due to the improved quality prediction of signals with coding errors caused by music signals that are coded with bandwidth extension techniques causing small impairments. Motivated by these results, future work should -- for example -- investigate further phenomena such as co-modulation \cite{Par2056_2019} and other perceptual organization models \cite{bregman1994auditory}, and their interaction with different distortion metrics in PEAQ or other multidimensional PQMSs.

\clearpage
\bibliographystyle{IEEEbib}
\bibliography{citation_list}

\end{document}